\documentclass{elsart}
\usepackage{graphicx}
\usepackage{amssymb}
\usepackage{wasysym}
\usepackage{latexsym,textcomp} 

\begin{document}
\begin{frontmatter}

\title{Planar LAAPDs: Temperature Dependence, Performance, and
Application in Low Energy X--ray Spectroscopy}

\author[unifr,psi,cor1]{L.~Ludhova}
\author[coim]{F.D.~Amaro},
\author[mpq]{A.~Antognini},
\author[lkb]{F.~Biraben},
\author[coim]{J.M.R.~Cardoso},
\author[coim]{C.A.N.~Conde},
\author[coim]{D.S.~Covita},
\author[yale,dax]{A.~Dax},
\author[yale]{S.~Dhawan},
\author[coim]{L.M.P.~Fernandes},
\author[mpq]{T.W.~H\"{a}nsch},
\author[yale]{V.-W.~Hughes},
\author[unifr]{O.~Huot},
\author[lkb]{P.~Indelicato},
\author[lkb]{L.~Julien},
\author[unifr]{P.E.~Knowles},
\author[eth]{F.~Kottmann},
\author[coim]{J.A.M.~Lopes},
\author[taiw]{Y.-W.~Liu},
\author[coim]{C.M.B.~Monteiro},
\author[unifr,fm]{F.~Mulhauser},
\author[lkb]{F.~Nez},
\author[psi,mpq]{R.~Pohl},
\author[prince]{P.~Rabinowitz},
\author[coim]{J.M.F.~dos~Santos},
\author[unifr]{L.A.~Schaller},
\author[psi]{D.~Taqqu},
\author[coim]{J.F.C.A.~Veloso}

\corauth[cor1]{Corresponding author. Tel.: +41 56 3103758; Fax: +41 56
  310 5230. {\em E-mail address:} Livia.Ludhova@psi.ch}
\address[unifr]{D\'epartement de Physique, Universit\'e de Fribourg, CH--1700 Fribourg, Switzerland}
\address[psi]{Paul Scherrer Institut, CH--5232 Villigen PSI, Switzerland}
\address[coim]{Departamento de F\'\i sica, Universidade de Coimbra, PT--3000, Coimbra, Portugal}
\address[mpq]{Max--Planck Institut f\"ur Quantenoptik, D--85748 Garching, Germany}
\address[lkb]{Laboratoire Kastler Brossel, \'Ecole Normale Sup\'erieure
  et Universit\'e P. et M. Curie, F--75252 Paris, CEDEX 05, France}
\address[yale]{Physics Department, Yale University, New Haven, CT 06520--8121, USA}
\address[eth]{Labor f\"ur Hochenergiephysik, ETH-H\"onggerberg, CH--8093 Z\"urich, Switzerland}
\address[taiw]{Physics Department, National Tsing Hua University, Hsinchu 300, Taiwan}
\address[prince]{Chemistry Department, Princeton University, Princeton, NJ 08544--1009, USA}
\thanks[dax]{Present address: CERN, CH--1211 Geneva 23, Switzerland}
\thanks[fm]{Present address: University of Illinois at Urbana--Champaign, IL 61801, USA}

\begin{abstract}
An experiment measuring the $2S$ Lamb shift in muonic hydrogen $(\mu ^-
p)$ is being performed at the Paul Scherrer Institute, Switzerland.
It requires small and compact detectors for 1.9~keV x~rays ($2P$--$1S$
transition) with an energy resolution around 25\% at 2~keV, a time
resolution better than 100~ns, a large solid angle coverage, and
insensitivity to a 5~T magnetic field.
We have chosen Large Area Avalanche Photodiodes (LAAPDs) from
Radiation Monitoring Devices as x--ray detectors, and they were used
during the last data taking period in 2003.
For x--ray spectroscopy applications, these LAAPDs have to be cooled
in order to suppress the dark current noise, hence, a series of tests
were performed to choose the optimal operation temperature.
Specifically, the temperature dependence of gain, energy resolution,
dark current, excess noise factor, and detector response linearity was
studied.
Finally, details of the LAAPDs application in the muonic hydrogen
experiment as well as their response to alpha particles are presented.
\end{abstract}

\end{frontmatter}

\section{Introduction}
\label{sec:Intro}

An experiment measuring the $2S$ Lamb shift $\Delta E(2P-2S)$ in muonic
hydrogen $(\mu ^- p)$ by precision laser spectroscopy is being
performed at the Paul Scherrer Institute (PSI),
Switzerland~\cite{kottm01}.
The experiment requires the detection of 1.9~keV
x--rays from the muonic hydrogen $K_{\alpha}$ Lyman line.
The apparatus is installed in a muon beam area at the PSI proton
accelerator, an environment with a rather high level of
neutron--induced radiative processes as well as electromagnetic and
acoustic noise.

The 1.9~keV x--ray detector has to reach an energy resolution of
$\sim$25\% and a time resolution better than 100~ns.
To optimize the solid angle for x--rays, the detector has 
to be mounted as near as possible to the pencil--shaped 
volume where the  $ \mu ^- p $  atoms are formed.
There is space for two sets of x--ray detectors (with sensitive areas
up to $\sim 2 \times 17\:\mbox{cm}^2$) at the top and bottom side of
the gas target which is mounted inside a solenoid with 20~cm inner
diameter.
The magnetic field of 5~T produced by the solenoid is another
limitation for the detector choice.
In addition, the whole target and detector setup is operated in
vacuum.

The experiment was installed for the first time in the
muon beam area during an engineering run in 2002.
A second beam period followed in 2003 during which the apparatus was
further improved.
Valuable data were taken in the last few weeks of the 2003 run with
the aim to search for the $2P-2S$ resonance.

\begin{figure}[htb]
\centerline{\includegraphics[width=0.5\linewidth]{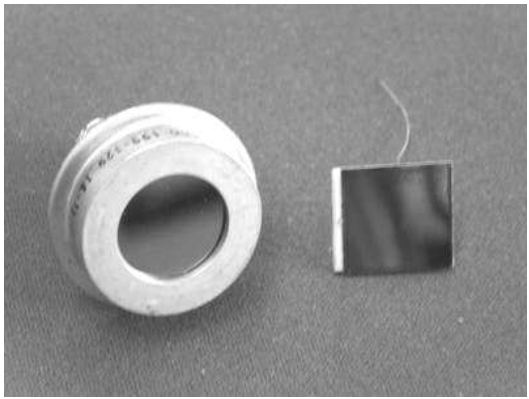}}
      \caption{\em Circular beveled--edge API~LAAPD (left) with
       a 16~mm diameter active surface area and square planar
       RMD~LAAPD (right) with $13.5\times 13.5\mbox{~mm}^{2}$
       active surface.}
\label{fig:laapd_rmd_api}
\end{figure}

In 2002 we used Large Area Avalanche Photodiodes from Advanced
Photonix Inc.~\cite{API} (API LAAPDs), a representative of the
beveled--edge LAAPDs, reviewed, for example, in Ref.~\cite{moszy02}.
They are circular, with a 16~mm diameter active surface surrounded by
a $\sim5$~mm wide ring of inactive material
(Fig.~\ref{fig:laapd_rmd_api}).
Their behavior in high magnetic fields was studied in
Refs.~\cite{ferna03,bouch03}, while a systematic investigation of
their low temperature performance in x--ray and visible--light
detection can be found in Ref.~\cite{ferna04}.
An example of their application to muonic atom spectroscopy during
the first stages of our experiment is given in Ref.~\cite{ferna03a}.

%1

During the most recent measurements in 2003 we replaced the API diodes
by LAAPDs from Radiation Monitoring Devices Inc.~\cite{RMD} (RMD
LAAPDs).
Two arrays of these detectors mounted below and above the pencil shaped
target have a considerably larger solid angle for x--ray detection due
to the LAAPD square shape and better ratio of sensitive to total
detector area (Fig.~\ref{fig:laapd_rmd_api}).
The principle of the RMD LAAPDs, in particular their
novel planar bevel fabrication process, is summarized in
Ref.~\cite{farre00}. 
In Sec.~\ref{sec:RMD} we describe these LAAPDs, their performance and
the results of systematic tests.
Section~\ref{sec:Application} contains some details about the RMD
LAAPDs application in our experiment, in particular their response to
alpha particles.
A comparison with the previously used API LAAPDs is given in the
conclusions.

\section{Properties of the RMD LAAPDs}
\label{sec:RMD}

\subsection{First tests}
\label{subsec:first}

\begin{figure}[htb]
\centerline{\includegraphics[angle=90,width=0.9\linewidth]{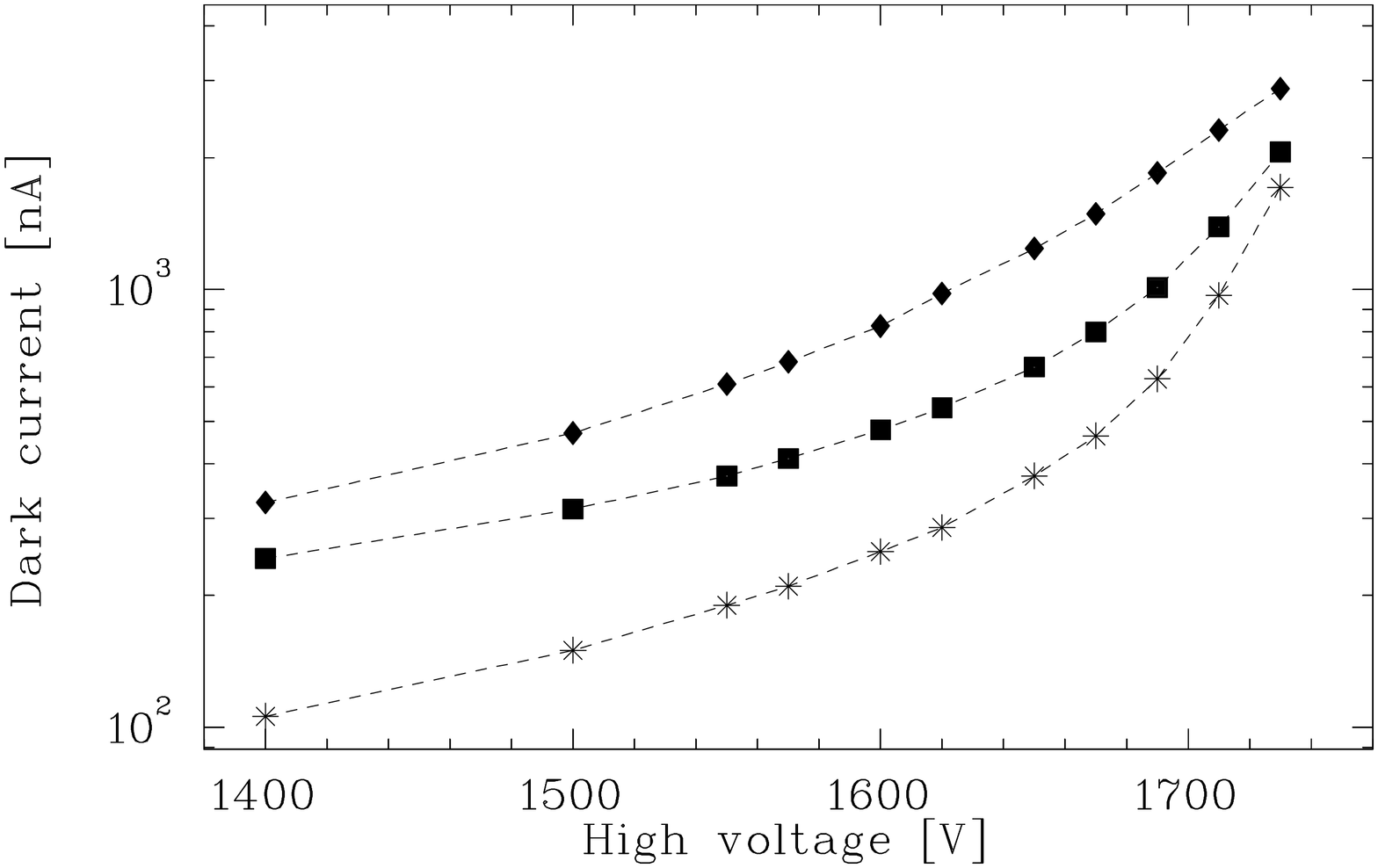}}
           \caption{\em LAAPD dark current versus high voltage measured at
           1\textcelsius\ (\textasteriskcentered), 10\textcelsius\
           ($\blacksquare$), and 23\textcelsius\ ($\blacklozenge$).}
\label{fig:I_HV}
\end{figure}

RMD LAAPDs, model S1315, are square planar devices with
13.5~mm$\times$13.5~mm sensitive area surrounded by 1~mm wide borders
of inactive material.
The operational voltage indicated by the manufacturer is in the region
from 1560 to 1765~V at 23\textcelsius\@.
The first tests with a $^{55}$Fe source emitting 5.9~keV x rays
performed at temperatures above 0\textcelsius\ have shown that these
LAAPDs require cooling to temperatures well below zero degrees Celsius
in order to detect low energy x~rays with an acceptable resolution.
The main reason is a high dark current reaching 1~$\mu$A at 1620~V at
room temperature; when cooled to 1\textcelsius, 1~$\mu$A is reached at
1710~V (Fig.~\ref{fig:I_HV}).

\begin{figure}[htb]
\centerline{\includegraphics[angle=90,width=0.9\linewidth]{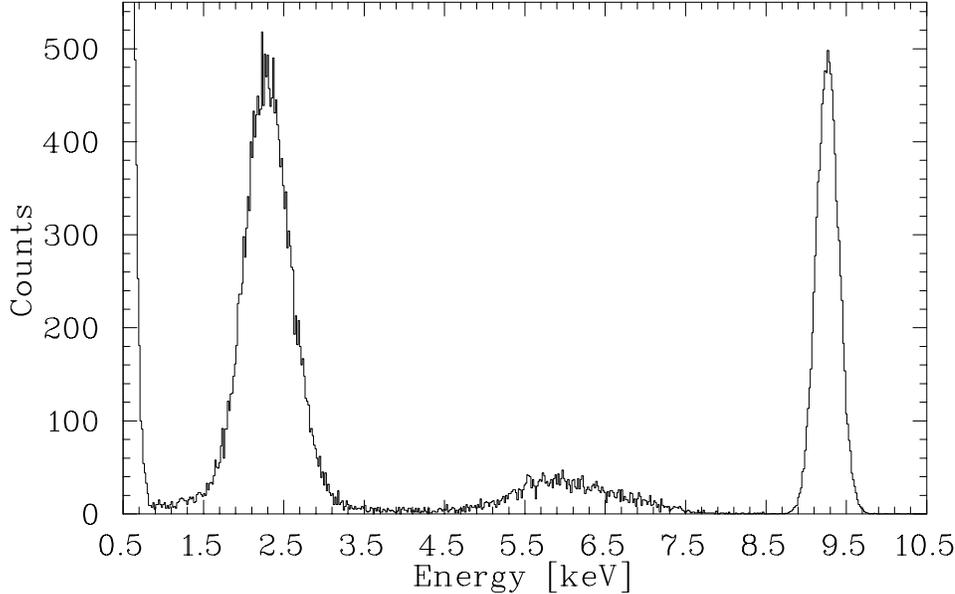}}
          \caption{\em LAAPD energy spectrum measured at $-23$\textcelsius\ with
          2.3~keV and 5.9~keV x--ray peaks and a peak due to the test
          pulses from a pulse generator.}
\label{fig:S_Fe_pulser}
\end{figure}

At room temperature, the noise tail in the energy spectrum extends up
to $\sim6$~keV\@.
By cooling below $-20$\textcelsius, satisfactory results were obtained
not only for 5.9~keV x~rays, but also for 2.3~keV x~rays, for which an
energy resolution of 26\% FWHM was reached.
The 2.3~keV x~rays were produced by sulfur fluorescing when irradiated
by a strong $^{55}$Fe source.
An example of such a spectrum measured at $-23$\textcelsius\ is shown
in Fig.~\ref{fig:S_Fe_pulser} where 2.3~keV and 5.9~keV x--ray peaks,
as well as a peak due to the test pulses from a pulse generator, which
are fed directly to the preamplifier, are visible.
The noise tail ends at 0.9~keV which makes x--ray spectroscopy around
2~keV well feasible.
%

%2
%

%3

\subsection{Experimental set--up}
\label{subsec:set_up}

In order to choose the optimal working temperature and to better
understand the temperature dependence of the RMD LAAPD performance
both for x--ray and visible--light detection, a series of tests were
performed.
X~rays with energies up to 15~keV from $^{54}$Mn, $^{55}$Fe, and
$^{57}$Co radioactive sources were used.
Light pulses from a light emitting diode (LED) with a peak emission at
635~nm were carried by a light guide to the LAAPD surface.
The LED light intensity was varied to achieve an energy equivalent
(LAAPD pulse height) of 11 to 14.5~keV\@.

The LAAPDs were operated in a light--tight, thermally insulated box.
A constant flow of cold nitrogen gas, regulated by a heater submerged
in a container of liquid nitrogen, was used for LAAPD cooling with a
temperature stabilization within $\pm~0.5$\textcelsius.

The LAAPD signals were fed into a low--noise charge--sensitive
integrating RAL~108A~preamplifier~\cite{ral} followed by a linear
amplifier HP~5582A, for which a gain between 16 and 128 as well as a
200~ns shaping time constant were used.
A PC--based, 1024--channel analyzer Nucleus PCA II was used to record
the energy spectra.

\subsection{Gain measurements}
\label{subsec:gain}

\begin{figure}[htb]
\centerline{\includegraphics[angle=90,width=0.9\linewidth]{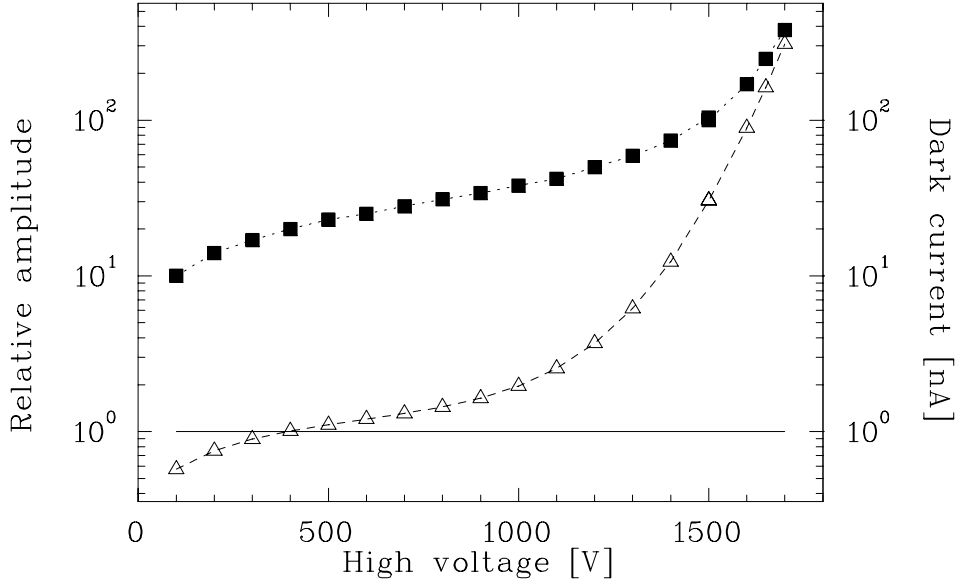}}
          \caption{\em LAAPD relative amplitude of the visible--light pulses
	  ($\triangle$) and the dark current ($\blacksquare$) versus
	  high voltage; measurement performed at 7\textcelsius. The
	  horizontal line represents the unity gain.}
\label{fig:unity_gain}
\end{figure}

Absolute gain measurements rely on the determination of unity gain,
which was found by averaging the amplitudes of 100~ns visible--light
LED pulses (635~nm wavelength) measured at a~bias voltage in the range
from 300 to 500~V\@.
For bias voltages below 300~V the recombination of the primary
electron--hole pairs plays an important role and the absolute gain is
below one.
Figure~\ref{fig:unity_gain} shows the relative amplitudes of the light
pulses as a function of high voltage, together with the dark current
observed during the measurement.
A horizontal line shows the unity gain. 

%4
\begin{figure}[htb]
\centerline{\includegraphics[angle=90,width=0.9\linewidth]{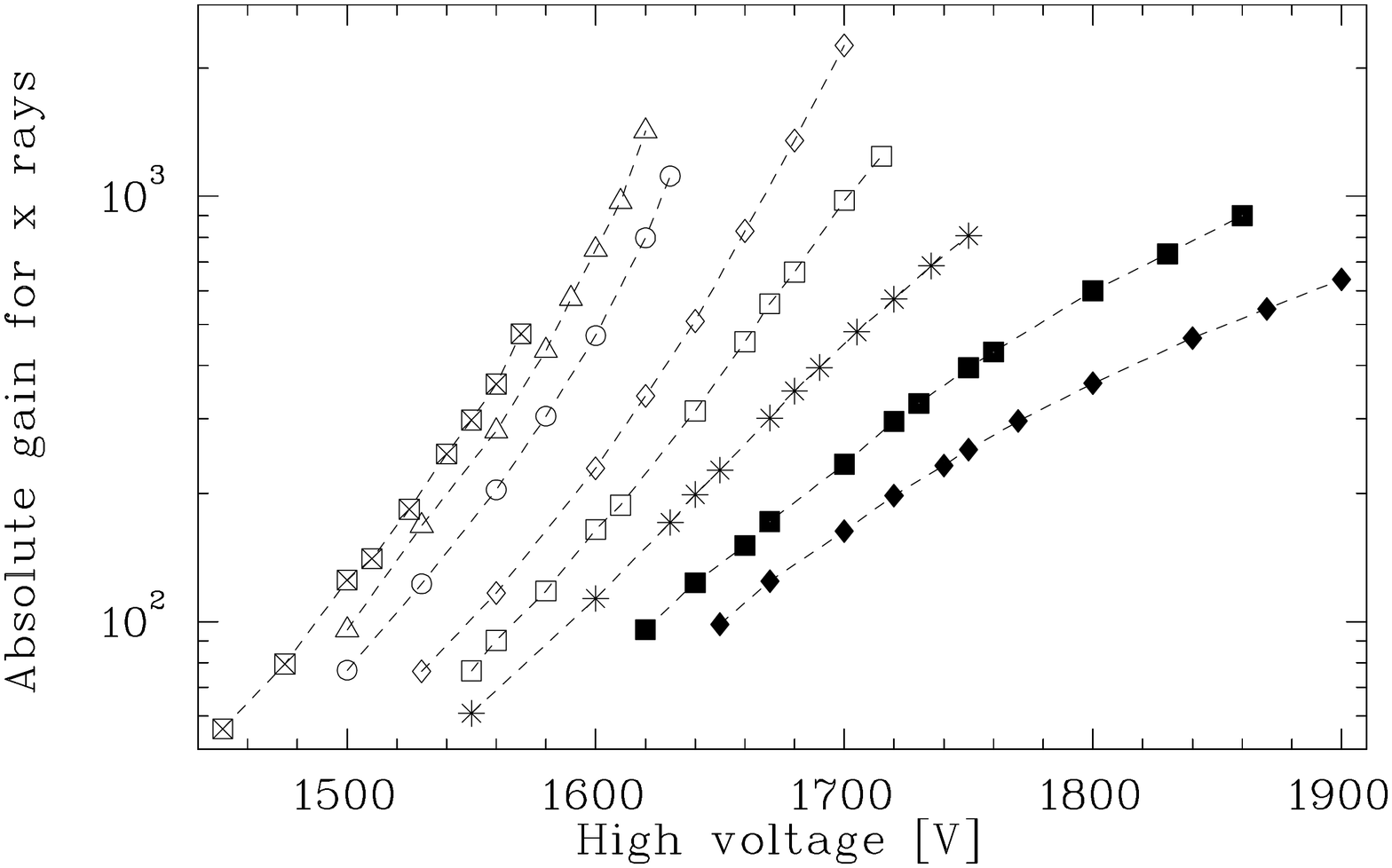}}
       \caption{\em LAAPD absolute x--ray gain versus high voltage measured
       at $-46$\textcelsius\ ($\boxtimes$), $-40$\textcelsius\
       ($\triangle$), $-33$\textcelsius\ ($\ocircle$),
       $-17$\textcelsius\ ($\Diamond$), $-8$\textcelsius\ ($\square$),
       0\textcelsius\ (\textasteriskcentered), 10\textcelsius\
       ($\blacksquare$), and 17\textcelsius\ ($\blacklozenge$).}
\label{fig:X_gain_HV}
\end{figure}

Assuming that the visible--light and the x--ray gains are equal at low
LAAPD gains, the absolute gain for x~rays was determined with a
$^{55}$Fe source emitting 5.9~keV x~rays.
Absolute x--ray gain versus high voltage, for the temperature range
from $-46$\textcelsius\ to 17\textcelsius, is given in
Fig.~\ref{fig:X_gain_HV}.

%5
\begin{figure}[htb]
\centerline{\includegraphics[angle=90,width=0.9\linewidth]{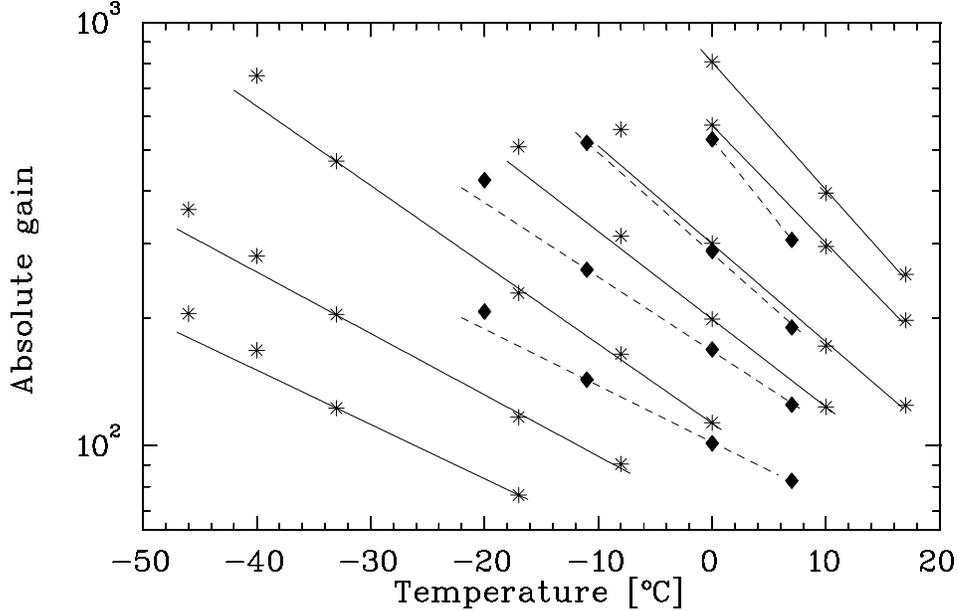}}
        \caption{\em LAAPD absolute x--ray (\textasteriskcentered) and
	visible--light ($\blacklozenge$) gain versus
	temperature. Measurements were performed at constant high
	voltages with x~rays (solid lines) at (from right to left)
	1530, 1560, 1600, 1640, 1670, 1720, and 1750~V and with
	visible light (dashed lines) at 1590, 1625, 1660, and
	1700~V\@.}
\label{fig:X_light_gain_T}
\end{figure}

For a given bias voltage, the gain increases exponentially with
decreasing temperature as seen in Fig.~\ref{fig:X_light_gain_T}.
The dependency is more pronounced for higher bias voltages and similar
for both x~rays (solid lines) and visible light (dashed lines).
Below a certain temperature the gain starts to increase even more
rapidly, as it is demonstrated in Fig.~\ref{fig:X_light_gain_T}.
%

%6
\subsection{Energy resolution}
\label{subsec:resolution}

\begin{figure}[hbt]
\centerline{\includegraphics[angle=90,width=0.9\linewidth]{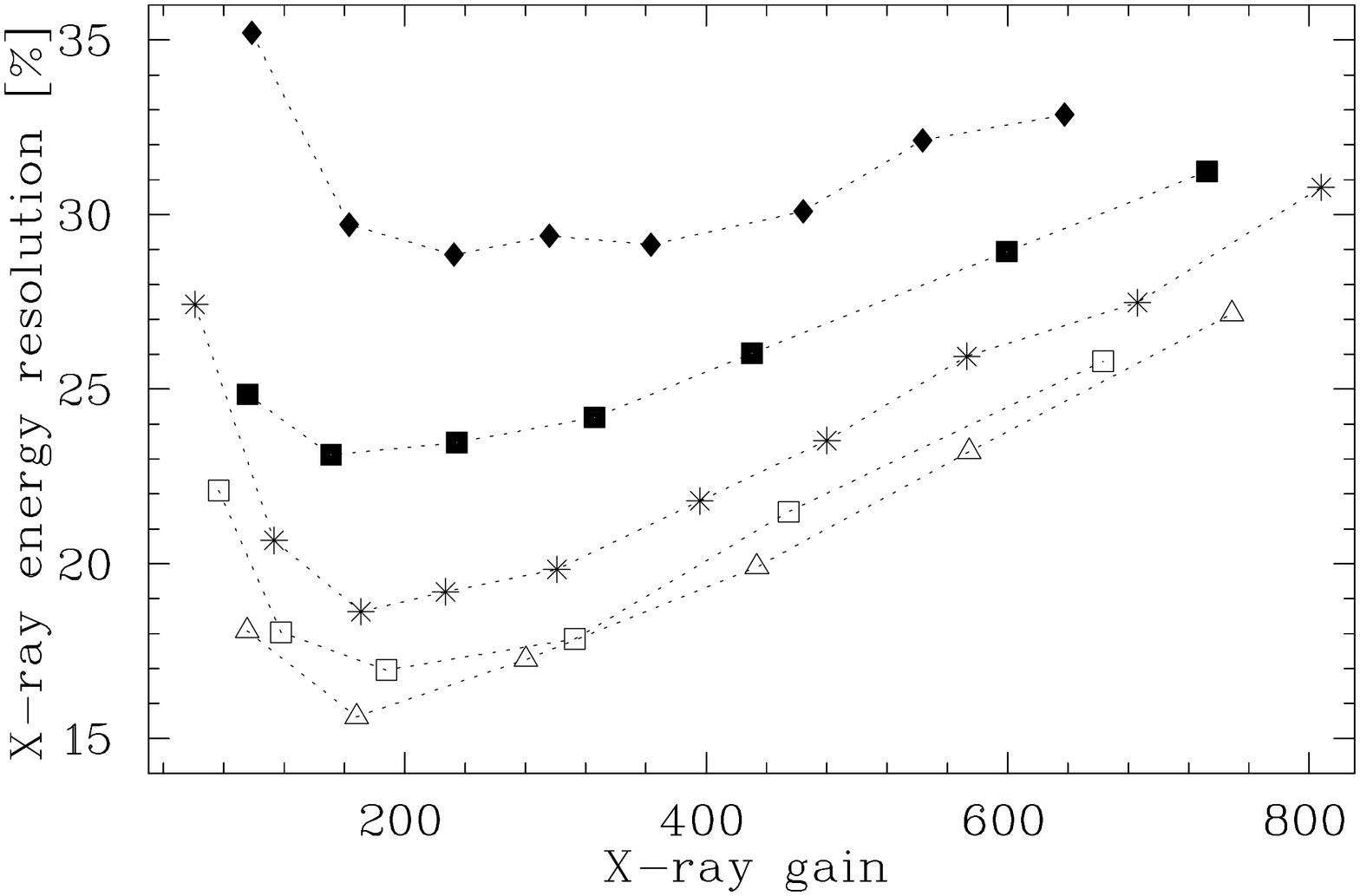}}
      \caption{\em LAAPD energy resolution for 5.9~keV x~rays versus gain
      measured at $-40$\textcelsius\ ($\triangle$), $-8$\textcelsius\
      ($\square$), 0\textcelsius\ (\textasteriskcentered),
      10\textcelsius\ ($\blacksquare$), and 17\textcelsius\
      ($\blacklozenge$).}
\label{fig:X_res_gain}
\end{figure}

\begin{figure}[htb]
\centerline{\includegraphics[angle=90,width=0.9\linewidth]{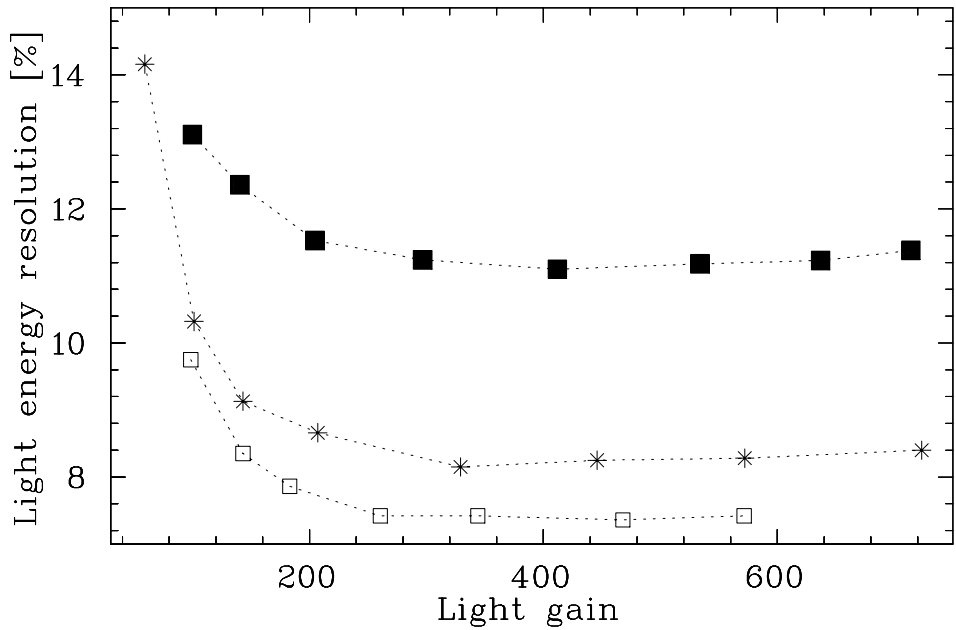}}
       \caption{\em LAAPD energy resolution for visible light versus
       gain measured at $-12$\textcelsius\ ($\square$), 0\textcelsius\
       (\textasteriskcentered), and 10\textcelsius\
       ($\blacksquare$). The energy equivalent of the light pulses
       corresponds to 11--14.5~keV x rays.}
\label{fig:L_res_gain}
\end{figure}

At low gain values the energy resolution improves with increasing
gain.
This trend continues up to a gain around 200 where the optimum is
obtained, for both 5.9~keV x~rays (Fig.~\ref{fig:X_res_gain}) and
visible light (Fig.~\ref{fig:L_res_gain}).
This optimal gain value does not depend on the temperature.
Higher gain increases the effect of spatial nonuniformity of the LAAPD
gain.
Due to the local character of the x--ray interaction with an LAAPD,
this effect worsens the x--ray energy resolution.
For light detection the whole illuminated area contributes to the
output signal, averaging local gain variations~\cite{moszy00}.
Consequently and in contrast to the x~rays, the visible--light energy
resolution remains constant at gains above 250.
%

%7
%

%8
\begin{figure}[hbt]
\centerline{\includegraphics[angle=90,width=0.9\linewidth]{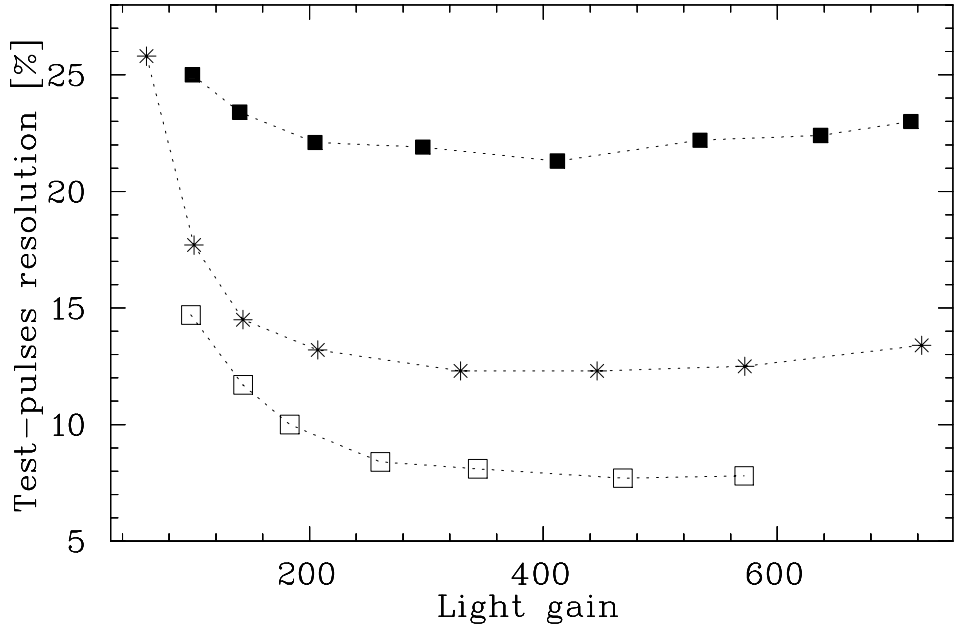}}
       \caption{\em LAAPD energy resolution for the test pulses versus gain,
       normalized to 5.9~keV. Measurements at $-12$\textcelsius\
       ($\square$), 0\textcelsius\ (\textasteriskcentered), and
       10\textcelsius\ ($\blacksquare$).}
\label{fig:N_res_gain}
\end{figure}

Simultaneous measurements of 5.4~keV x~rays, visible--light pulses, and
test signals from a pulse generator were made at different gains and
temperatures.
The energy calibration was determined by the x--ray peak and the LAAPD
gain was deduced from the position of the visible--light peak.
The position of the test--pulse peak does not depend on the LAAPD
gain.
Its width represents the LAAPD dark current and preamplifier noise
contributions to the overall resolution~\cite{moszy02}.
To express the test--pulse width as a relative energy resolution, it
has to be normalized to a given energy which was chosen to be 5.9~keV,
as shown in Fig.~\ref{fig:N_res_gain}.
Also here, the resolution reaches the minimum at a gain of about 200 and
stabilizes for higher gains.
The similar behavior of the visible--light and test--pulse resolutions
is explained in Sec.~\ref{subsec:ENF}.
%

%9
\subsection{Dark current}
\label{subsec:dark}

\begin{figure}[htb]
\centerline{\includegraphics[angle=90,width=0.9\linewidth]{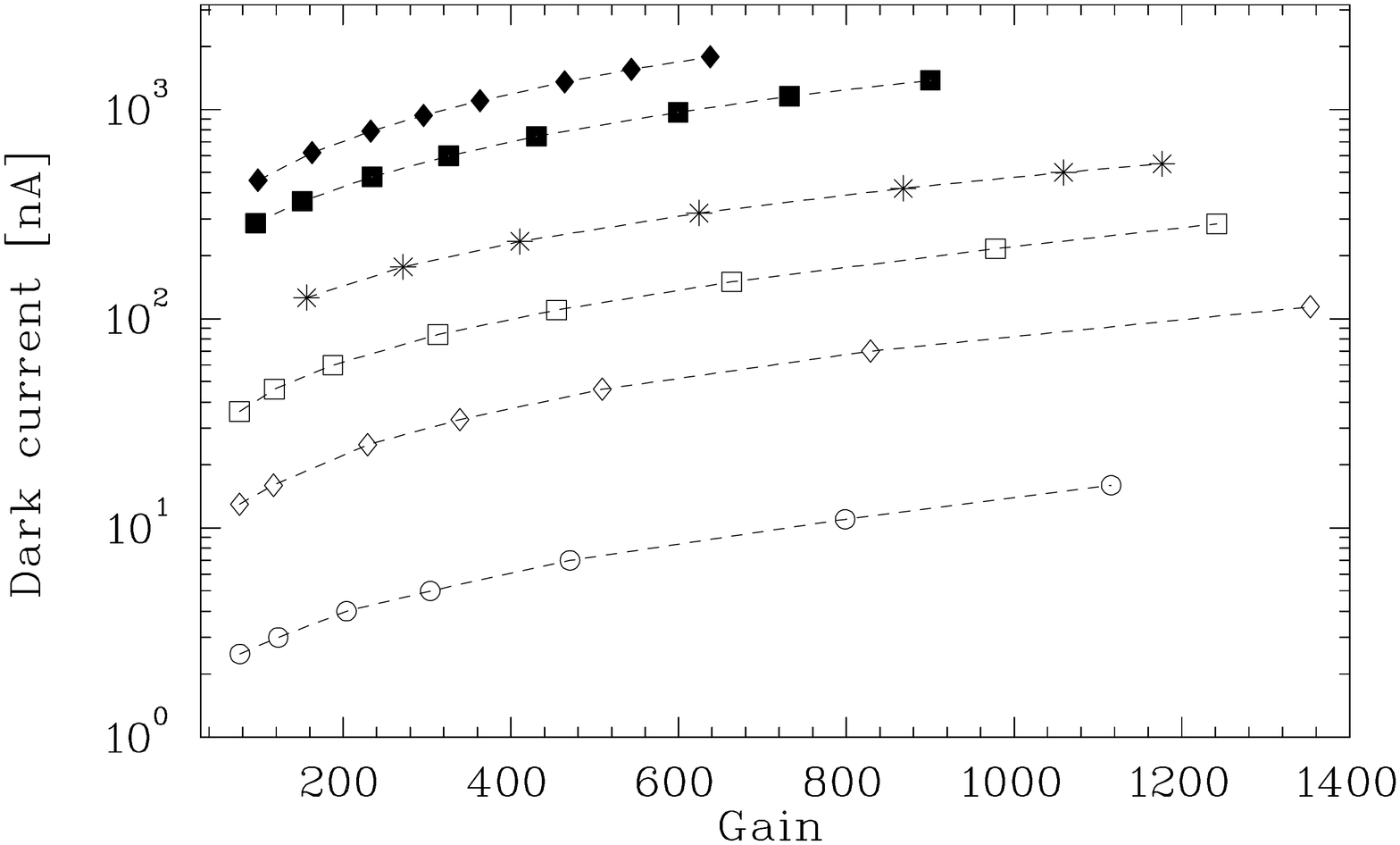}}
        \caption{\em LAAPD dark current versus gain measured at
        $-33$\textcelsius\ ($\ocircle$), $-17$\textcelsius\
        ($\Diamond$), $-8$\textcelsius\ ($\square$), 0\textcelsius\
        (\textasteriskcentered), 10\textcelsius\ ($\blacksquare$), and
        17\textcelsius\ ($\blacklozenge$).}
\label{fig:current_gain}
\end{figure}

The dark current depends strongly on the temperature and LAAPD gain.
At a given temperature it increases with gain and is reduced by an
order of magnitude for each 20\textcelsius\ temperature decrease, as
shown in Fig.~\ref{fig:current_gain}.
At $-33$\textcelsius\ and in the gain region below 800 the dark
current is below 10~nA\@.
%

%10
\subsection{Excess noise factor}
\label{subsec:ENF}

Measurements of the 5.4~keV x--rays from a $^{54}$Mn source,
visible--light, and test pulses were made simultaneously for different
temperatures and different gains, allowing us to study the temperature
dependence of the excess noise factor $(F)$, defined~\cite{moszy02} as
\begin{equation}
 F=1+\sigma_{A}^{2}/M^{2} \, ,
\end{equation}
where $M$ is the LAAPD gain and $\sigma_{A}$ its variance.
Neglecting the small light--intensity fluctuations and considering
that detector inhomogeneity contributions to the energy resolution are
averaged out for the light measurements, the energy resolution $\Delta
E$ (FWHM) of the light peak with a Gaussian shape can be
described~\cite{moszy02} as
\begin{equation}
\Delta
E^{2}=(2.355)^{2}FE\varepsilon+\Delta ^{2}_{\rm noise}
\label{eq:resolution}
\end{equation}
where $E$ is the energy equivalent of the light peak, $\varepsilon$ is
the energy per electron--hole pair creation in Si (3.6~eV), and
$\Delta^{2}_{\rm noise}$ is the dark noise contribution of the
diode--preamplifier system.

\begin{figure}[hbt]
\centerline{\includegraphics[angle=90,width=0.9\linewidth]{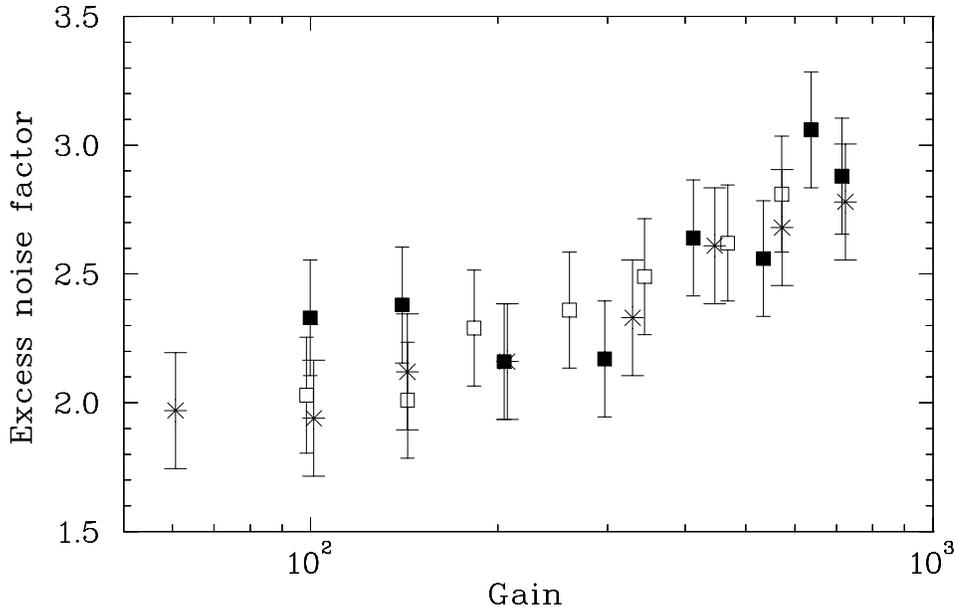}}
      \caption{\em LAAPD excess noise factor versus gain measured at
      $-12$\textcelsius\ ($\square$), 0\textcelsius\
      (\textasteriskcentered), and 10\textcelsius\ ($\blacksquare$).}
\label{fig:ENF_gain}
\end{figure}

In the energy spectrum, the x--ray peak gives the energy calibration.
The FWHM of the test pulses peak defines $\Delta^{2}_{\rm noise}$.
The deduced value of $F$ was found to be temperature independent and
slowly increasing with the LAAPD gain (Fig.~\ref{fig:ENF_gain}).
A faster increase at gains above 300 reflects the contribution of
holes to the amplification process~\cite{moszy00}.
Typical values were $F\approx2.2$ at gain 200 and $F\approx2.8$ at
gain 600; this represents a 27\% increase.

As given by Eq.~(\ref{eq:resolution}), for light pulses with an energy
equivalent $E$ there are two contributions to the resolution $\Delta
E$.
By cooling, the contribution from the dark current noise
$\Delta^{2}_{\rm noise}$ is suppressed (Fig.~\ref{fig:N_res_gain}) and
the contribution due to the temperature independent increase of the
excess noise factor becomes relatively more significant.
However, it is important to note that a 27\% increase of $F$ is
accompanied by only a small increase of $\Delta E$ even at low
temperatures.
For the light pulses with an energy equivalent in the range
$11-14.5$~keV, the corresponding $\Delta E$ increase is below 4\%,
7\%, and 10\% at temperatures 10\textcelsius, 0\textcelsius, and
$-12$\textcelsius, respectively.
This also explains the similar behavior of the visible--light and
test--pulse resolutions as functions of LAAPD gain, as was shown
in Sec.~\ref{subsec:resolution} and in Figs.~\ref{fig:L_res_gain}
and~\ref{fig:N_res_gain}.
%

%11

\subsection{Nonlinearity}
\label{subsec:nonlin}

The use of the x--ray peak for the energy calibration of the light
peak is correct only if the LAAPD response is perfectly linear, i.e.,
if the resulting signal amplitude is proportional to the initial
number of electron--hole pairs.
In reality, there is a well--known nonproportionality between the
gains for x--ray and visible--light events, as well as between x--ray
events with different
energies~\cite{moszy02,ferna04,moszy00,pansa97,moszy03}.
In contrast to visible light, an x~ray interacting in the LAAPD
produces high charge densities causing both a decrease of the local
electric field and local heating.
The effect is important at higher gains and causes the x--ray gain
to be smaller than the visible--light gain.

\begin{figure}[htb]
\centerline{\includegraphics[angle=90,width=0.9\linewidth]{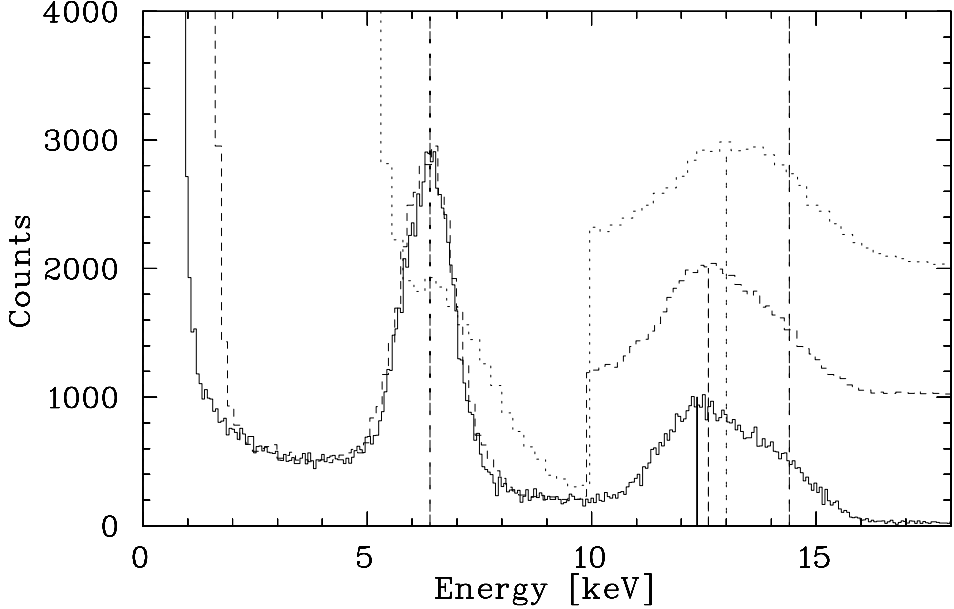}}
           \caption{\em LAAPD energy spectra from measurements of
           6.4~keV x~rays and 14.4~keV~$\gamma$~rays from a $^{57}$Co
           source, performed at $-24$\textcelsius\ (solid line),
           1\textcelsius\ (dashed line), and 27\textcelsius\ (dotted
           line) at gains 400, 350, and 200, respectively. Each
           spectrum was separately calibrated in energy such that the
           6.4~keV peak occurs at the same position. The maxima of the
           14.4~keV peaks are then at 12.4, 12.6, and 13.0~keV,
           respectively, due to the nonlinear LAAPD response. The
           parts of the spectra above 10~keV for the measurements at
           1\textcelsius\ and 27\textcelsius\ are shifted in the y
           direction by 1000 and 2000, respectively.}
\label{fig:co_spectrum}
\end{figure}

The nonlinearity for x~rays with different energies was measured with
a $^{57}$Co source by comparing the relative positions of the 6.4~keV
Fe~$K_{\alpha}$~line and the 14.4~keV $\gamma$ line.
A comparison of three such spectra measured at different temperatures
and different gains is presented in Fig.~\ref{fig:co_spectrum}.
%

%12
\begin{figure}[hbt]
\centerline{\includegraphics[angle=90,width=0.9\linewidth]{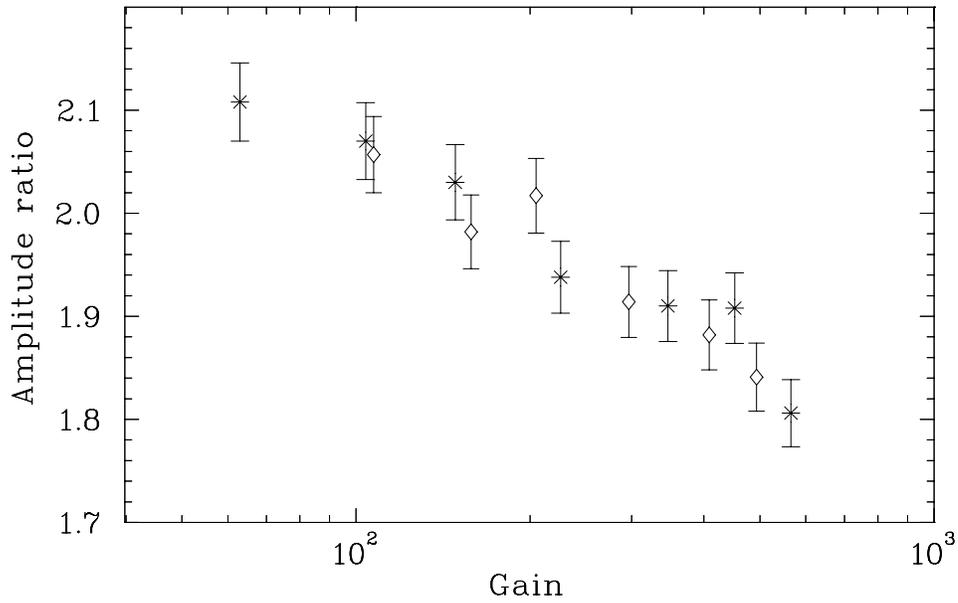}}
           \caption{\em Ratio of the amplitudes of 14.4 and 6.4~keV
	   signals versus LAAPD gain, measured at $-20$\textcelsius\
	   ($\Diamond$) and 1\textcelsius\ (\textasteriskcentered).}
\label{fig:linearity_x}
\end{figure}

The amplitude ratio of the 14.4 and 6.4~keV x--ray signals versus
gain, measured at $-20$\textcelsius\ and 1\textcelsius\ is shown in
Fig.~\ref{fig:linearity_x}.
The LAAPD nonlinear response for x~rays with different energies is
temperature independent; normalized to the value for linear operation,
$14.4/6.4$, the nonlinear effect is 11\% at gain 200 and 16\% at
gain 400.
%

%13
\begin{figure}[htb]
\centerline{\includegraphics[angle=90,width=0.9\linewidth]{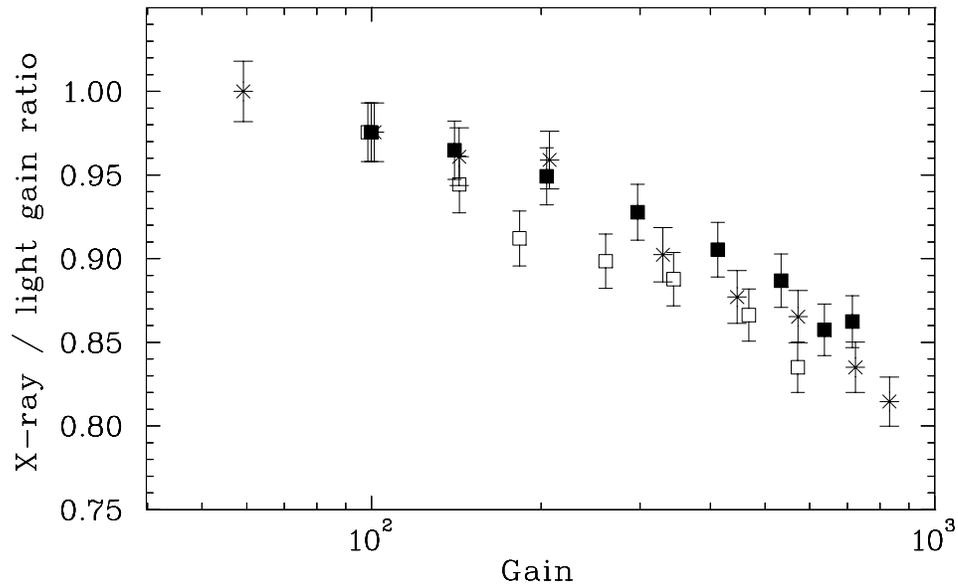}}
        \caption{\em Ratio of 5.4~keV x--ray to visible--light gains
        versus LAAPD gain, measured at $-12$\textcelsius\ ($\square$),
        1\textcelsius\ (\textasteriskcentered), and 10\textcelsius\
        ($\blacksquare$).}
\label{fig:linearity_XtoL}
\end{figure}

The ratio of the x--ray to visible--light gains, shown in
Fig.~\ref{fig:linearity_XtoL}, was measured by a simultaneous
illumination of the LAAPD by 5.4~keV x~rays and by visible--light
pulses.
A possible temperature dependence of this ratio, as was observed for
the API LAAPDs~\cite{ferna04}, is not visible given our measurement
uncertainty.
The nonlinearity effect is 5\% at a gain of 200, and reaches 10\% at a
gain of 400.
Assuming that the LAAPD response to the light pulses with an energy
equivalent of $11-14.5$~keV is linear, one can consider the
nonlinearity as an absolute nonlinearity for 5.4~keV x~rays.

Assuming that the nonlinearities for 5.4 and 6.4~keV x~rays are
similar, the absolute nonlinearity for 14.4~keV x~rays can be
estimated from the measurements shown in Figs.~\ref{fig:linearity_x}
and~\ref{fig:linearity_XtoL}.
It amounts to about 15\% at a gain of 200 and to about 24\% at a gain
of 400.

The high local charge density created in the LAAPD by an interacting
x~ray --- the reason for the nonlinear LAAPD response --- is
proportional to the number of electron--hole pairs, and hence, at a
given LAAPD gain, to the x--ray energy.
It is thus reasonable to assume that the nonlinearity at a
certain gain is, in first order, proportional to the x--ray energy.
The ratio of the nonlinearities for 14.4~keV x~rays to 5.4(6.4)~keV
x~rays is 3.0 and 2.4 for gains 200 and 400, respectively.
These ratios differ from the direct ratio of the energies
$14.4/5.4(6.4)=2.7(2.3)$ by less than 15\%, an error introduced by the
assumption of the same nonlinearity for 5.4 and 6.4~keV x~rays.
%

%14

\section{LAAPD application in the muonic $2S$ Lamb shift experiment}
\label{sec:Application}

\subsection{LAAPD operation conditions}
\label{subsec:operation}

\begin{figure}[htb]
\centerline{\includegraphics[width=0.5\linewidth]{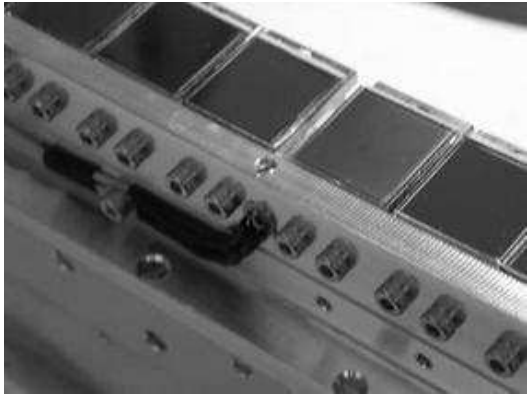}}
      \caption{\em Central part of one RMD LAAPD array. The wires are
       the thermometer leads electrical connections. High voltage
       connections and preamplifiers are located on the backside of
       each LAAPD.}
\label{fig:laapd_photo}
\end{figure}

During the most recent data taking period of the muonic $2S$ Lamb shift
experiment in 2003, two face--to--face rows of ten~RMD LAAPDs each
were mounted around the target.
Figure~\ref{fig:laapd_photo} shows the central part of one row.
The detector assembly was operated in a vacuum of $10^{-6}$~mbar and a
magnetic field of 5~Tesla.
%

%15
For an optimal measurement of the 1.9~keV x--ray line, we cooled the
whole mount to $-30$\textcelsius\@ by circulating cold methanol
through a small heat exchanger which was in thermal contact with the
LAAPDs.
The resolutions (FWHM) obtained for 5.9~keV x~rays varied for the
20~LAAPDs between 11\% and 20\%, with an average of about 15\%.
A tendency that LAAPDs with higher gain at a given high voltage have
better resolution was observed.

The typical bias voltage was around 1600~V and the corresponding gain
about 400, a value chosen for each LAAPD so that the amplitude of the
1.9~keV x--ray signal was sufficiently above the noise level without
worsening the resolution.
Based on the discussion in Sec.~\ref{subsec:nonlin} the nonlinearity
for x~rays with energies $\sim 2$~keV can be estimated to be about
$3-4$\% at gain 400.
The dark current was between $8-25$~nA for the majority of the LAAPDs.

\begin{figure}[hbt]
\centerline{\includegraphics[angle=90,width=0.9\linewidth]{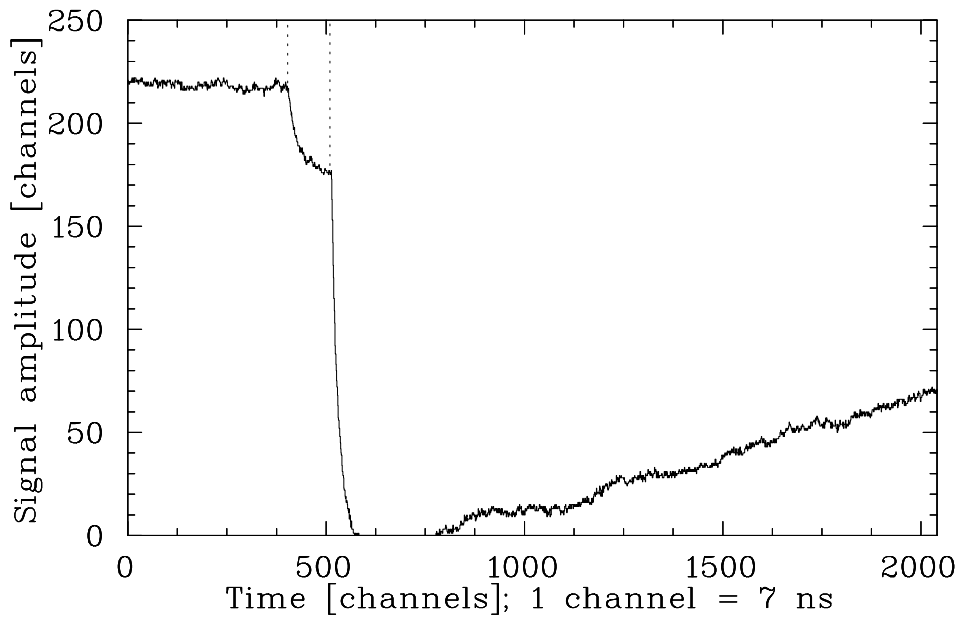}}
       \caption{\em A typical event with two superimposed LAAPD pulses
       ($\Delta t = 750$~ns) recorded by a wave--form digitizer. The
       pulses have negative amplitude. The small first pulse is a
       1.9~keV x--ray signal and the second larger pulse is due to a
       muon--decay electron. The digitizer dynamic range is between 0
       and 255, and hence, the second pulse is saturated between
       channels 600 and 800 and only gives 0 as reading. The beginning
       of each signal is marked by a vertical dotted line.}
 \label{fig:fadc}
\end{figure}

After the preamplifiers, the amplitude of a 1.9~keV x--ray signal was
about 2~mV.
The signal rise time for 25\% of the detectors was below $\sim250$~ns,
for 50\% was in the interval 300 to 450~ns and for 25\% was above
450~ns.
After further amplification, the negative amplitude signals were
stored in an 8--bit wave--form digitizer~\cite{faDC} operated at
140~MHz, which allowed an optimal suppression of background signals
with nonstandard shape and, in particular, permitted the separation of
two consecutive, superimposed pulses.
A typical event from one LAAPD is given in Fig.~\ref{fig:fadc}.
The baseline noise fluctuations, although small, cannot be neglected in
comparison with the amplitude of the 2~keV x~ray.
%

%16

\subsection{LAAPD response to alpha particles}
\label{subsec:alphas}

In our experiment, the LAAPDs were exposed to alpha particles with
kinetic energies from 2 to 9~MeV at a~rate of about
$5\:\mbox{s}^{-1}$.
The alpha particles came from the dielectric coating of two
high--reflectivity laser mirrors for 6~$\mu$m light which contain
thorium.
The mirrors were mounted only 1.5~cm away from the LAAPD surface.

\begin{figure}[hbt]
\begin{minipage}[t]{.48\textwidth}
\centerline{\includegraphics[angle=90,width=0.9\linewidth]{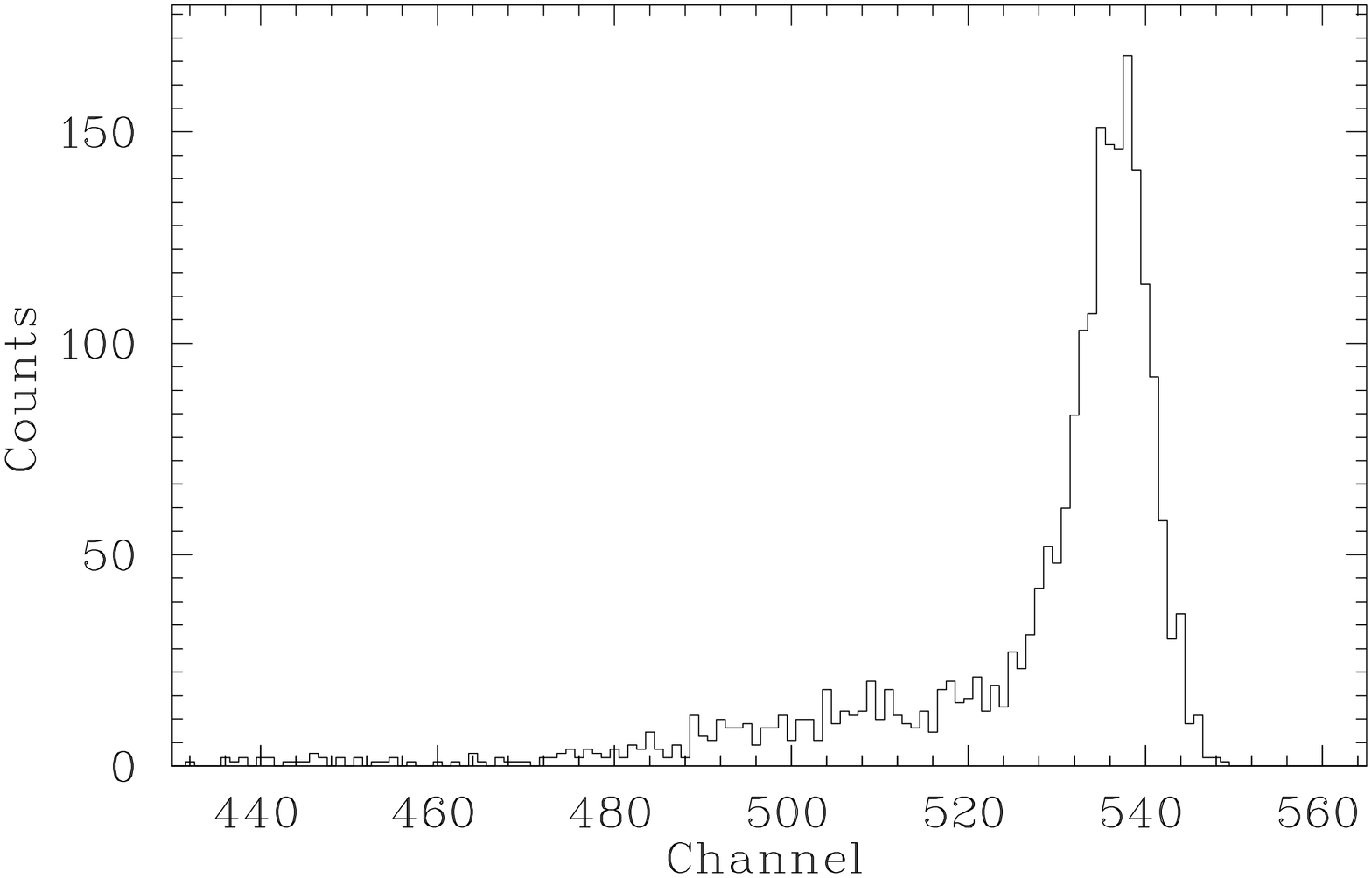}}
        \caption{\em LAAPD energy spectrum of the alpha particles from
        a $^{241}$Am source measured at 600~V\@. Due to energy loss in
        the source, the spectrum has a low energy tail.}
\label{fig:alpha-600}
\end{minipage}
\hfil
\begin{minipage}[t]{.48\textwidth}
\centerline{\includegraphics[angle=90,width=.9\linewidth]{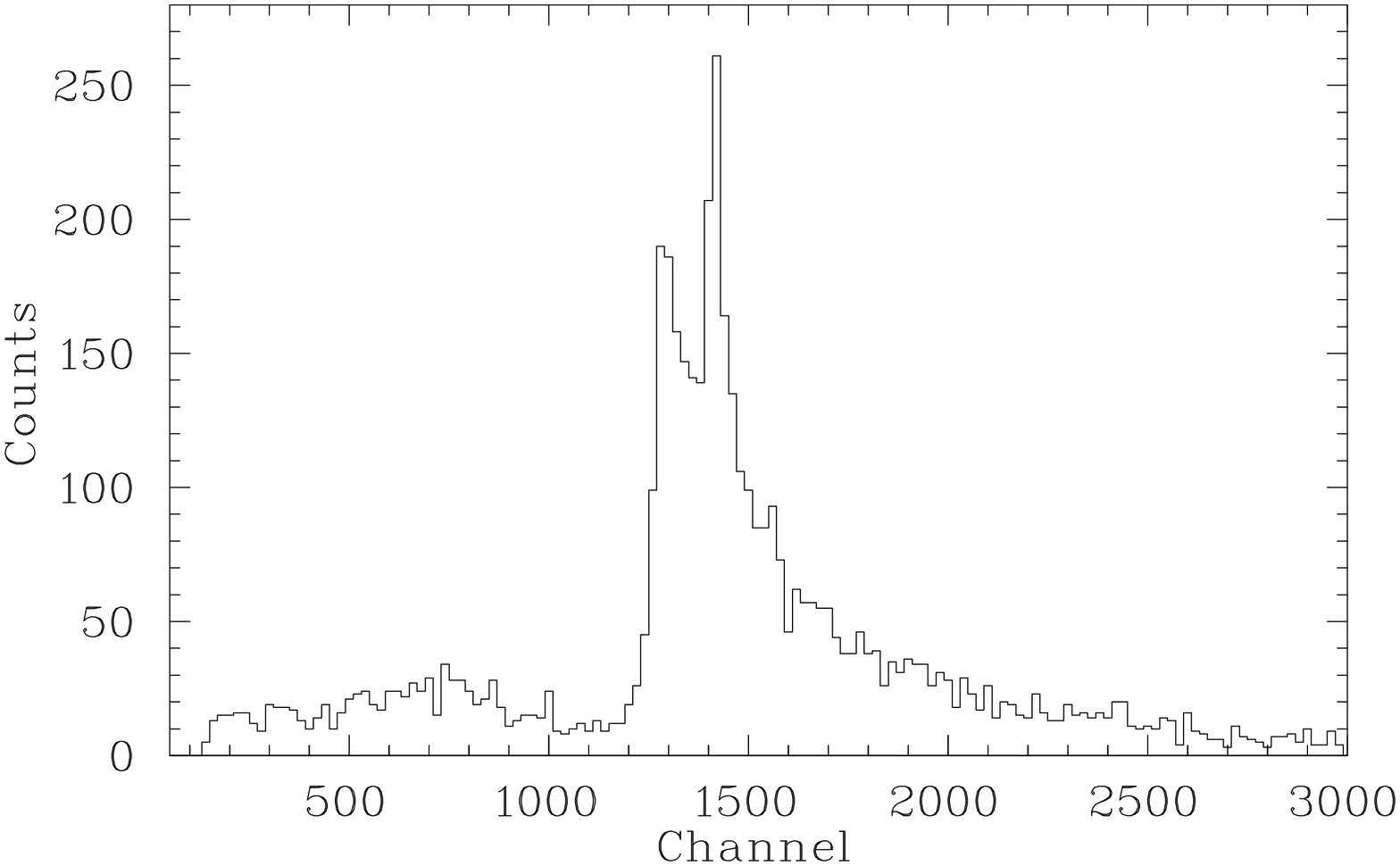}}
        \caption{\em LAAPD energy spectrum of the alpha particles as
        in Fig.~\ref{fig:alpha-600}, but measured at 1300~V\@. It
        corresponds to not saturated signals. Note the high--energy
        tail which is absent in Fig.~\ref{fig:alpha-600}. }
\label{fig:alpha-1300nice}
\end{minipage}
\end{figure}

The response of the API LAAPDs to alpha particles was studied with a
collimated $^{241}$Am alpha source (5.4~MeV) providing events at about
$20\:\mbox{s}^{-1}$.
Due to the high ionization density of alpha particles, they produce
signals measurable in the LAAPD even at room temperature and without
bias voltage.
The signal has a long rise time of about 500~ns and becomes
faster when bias voltage is applied.
At 600~V the rise time is about 230~ns and a typical alpha spectrum
with a low energy tail, due to energy losses in the source, is
observed (Fig.~\ref{fig:alpha-600}).
%

%17, 18

However, with a further high voltage increase a high energy tail
appears.
In addition, at about 1300~V, huge pulses are observed which
correspond to a LAAPD gain of order $10^{5}$.
The origin of these pulses is attributed to a plasma discharge in the
avalanche region along the very high ionization density of an alpha
particle trace~\cite{rmdpc}.
These pulses cause saturation of the preamplifier and have a long
recovery time.
In addition, the presence of these signals gives rise to large
fluctuations of the LAAPD dark current.
In a spectrum measured at 1300~V (Fig.~\ref{fig:alpha-1300nice}), one
may recognize a peak due to some alpha particles which are not
creating a plasma discharge. 
Not visible in Fig.~\ref{fig:alpha-1300nice} is the saturated signal
peak from the plasma discharge signals.

With an oscilloscope probe sensor we studied the shape of the plasma
discharge signals after the input coupling capacitor, while the
preamplifier was removed.
The LAAPD was operated at 1780~V\@.
The saturated signal has a long relaxation time of 200~ms and a huge
amplitude of 70~V\@.
For comparison, a 6~keV x~ray produces a signal of $\sim 10^{-4}$~V in amplitude.
Such a long recovery time represented an unacceptable dead time and the
high amplitude was dangerous for the preamplifier.
%

%19
\begin{figure}[hbt]
   \centerline{\includegraphics[width=0.9\linewidth]{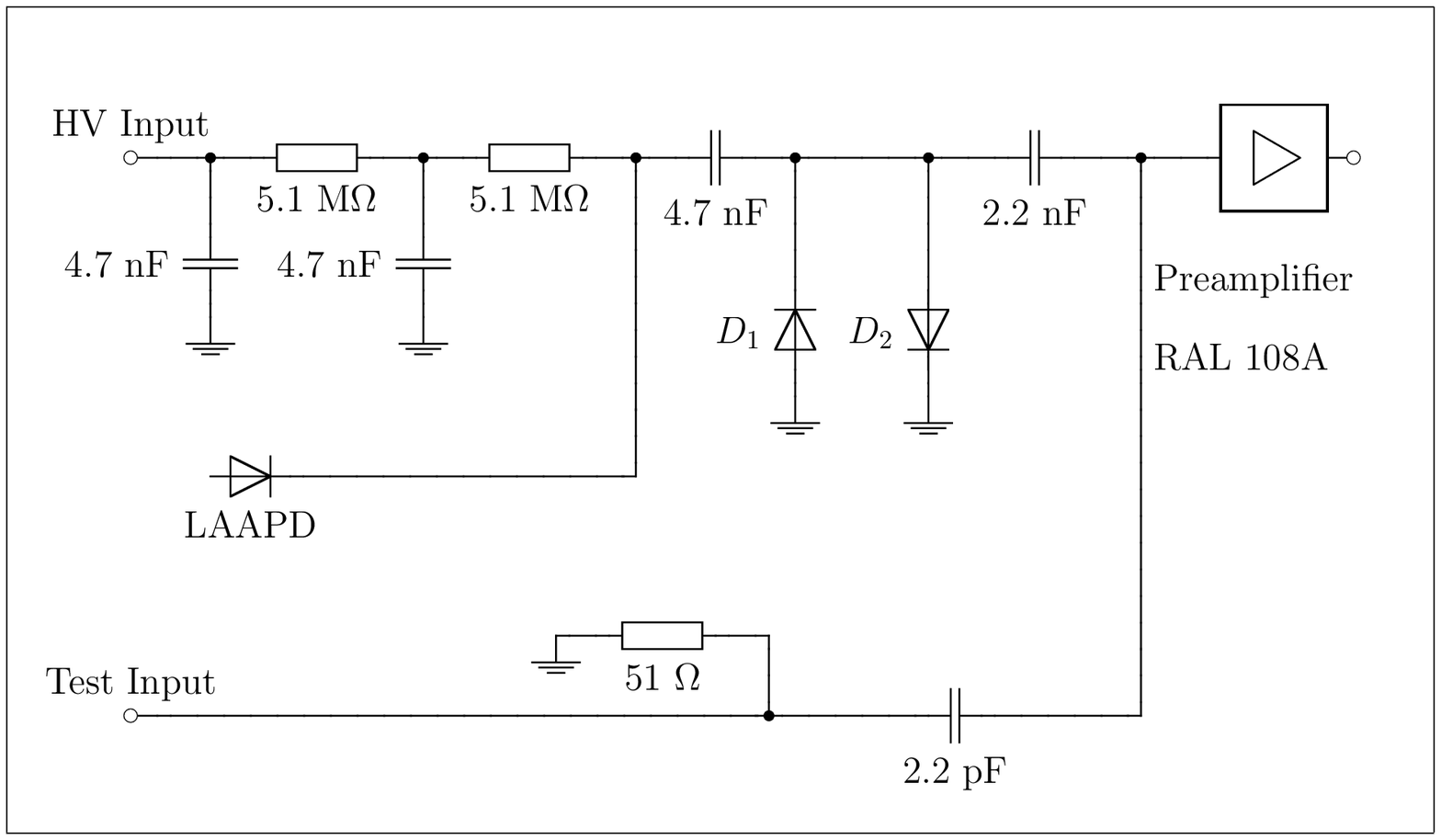}}
%   \vspace{-1cm}
   \caption{\em Final electronic scheme of a preamplifier box used for
       each RMD LAAPD. Note the two input coupling capacitors (4.7~nF
       and 2.2~nF) with the two diodes (D$_{1}$, D$_{2}$) in between;
       for explanation see text.}
\label{fig:preamp_box}
\end{figure}

Two high conductance ultra fast diodes 1N4150~\cite{fairchild}, chosen
for their fast recovery time of around 5~ns, were installed at the
input of the charge integrating RAL 108A~preamplifier~\cite{ral} to
limit the amplitude of plasma discharge signals.
To reduce the relaxation time, another input coupling capacitor was
added and placed after the two diodes.
Both input coupling capacitances were tuned in order to find an
optimum between shortening the relaxation time and losing the charge
sent to the preamplifier.
A good compromise was found by using a first capacitance of 4.7~nF and
a second of 2.2~nF\@.
The electronic scheme in Fig.~\ref{fig:preamp_box} represents the
final arrangement of the preamplifier electronics.
The recovery time was reduced by a factor 10 to reach 20~ms, measured
after the preamplifier.

At lower temperatures, the characteristics of plasma discharge signals
did not change.
The only difference was that they started to appear at lower voltages,
since at lower temperatures a certain gain is reached already at a
lower voltage.

In our application during the 2002 data taking period, after having
installed the laser mirrors and, hence, the alpha particles source,
three out of the ten API LAAPDs were destroyed within a day.
This happened after two weeks of perfect functioning, so there was a
high probability that the API LAAPDs were destroyed by the alpha
particles.

For the 2003 data taking period we used RMD LAAPDs as x--ray
detectors.
Due to their rectangular shape and to only a thin border of inactive
material, they covered a larger solid angle.
With 20~LAAPDs we were able to cover an average solid angle of 30\%
compared to 17\% obtained with the API LAAPDs.
Unfortunately, when exposed to alpha particles, their response was
very similar to that of the API LAAPDs.
To avoid the risk of LAAPD damage or slow deterioration, during the
final data taking we shielded the RMD LAAPDs with a 0.2~mm thick
lithium sheet, absorbing all alpha particles below 9~MeV, but
unavoidably also about 40\% of the 1.9 keV x~rays.

\section{Discussion and conclusions}
\label{sec:Conclusions}

The experiment measuring the $2S$ Lamb shift in muonic hydrogen,
performed at the Paul Scherrer Institute in Switzerland, has demanding
requirements for the 1.9~keV x--ray detectors.
The beveled--edge API LAAPDs used in the initial stages of the
experiment were replaced by the planar RMD LAAPDs for the latest data
taking.

In our experiment the LAAPDs are exposed to alpha particles causing
high amplitude signals with a long recovery time and, with a high
probability, detector damage or destruction.
No significant difference in the detector response to alpha particles
was found in between API and RMD LAAPDs.

The results of systematic tests studying the RMD LAAPDs performance at
low temperatures were shown.
In comparison, the API LAAPDs~\cite{ferna04} show in general better
performance and require less or no cooling, but the much higher solid
angle coverage achievable with the RMD LAAPDs is an essential
advantage for an application like ours.

In order to perform any reasonable soft x--ray spectroscopy
measurements, the RMD LAAPDs have to be cooled.
The dark current of RMD LAAPDs is of the order of a few $\mu$A at room
temperature whereas values around 10~nA are reached at
$-30$\textcelsius\@.
In contrast, the dark current of the API LAAPDs is of the order of a
few hundred nA at a room temperature and can be reduced to 10~nA
already at 0\textcelsius\@.

The RMD LAAPDs we used demonstrated worse energy resolution in
comparison with the API LAAPDs.
An 8\% energy resolution for visible light with the energy equivalent
of about 10~keV was obtained at room temperature with the API LAAPDs;
to reach the same resolution with the RMD LAAPDs, they needed to be
cooled to $-10$\textcelsius\@.
With the API LAAPDs an energy resolution of 11\% for 5.4~keV x~rays at
room temperature was measured; with the RMD LAAPDs the resolution of
11\% for 5.9~keV x~rays was reached only at $-30$\textcelsius\ with
the best LAAPDs.

The API LAAPDs operate with their optimal resolution at gain 50.
At this gain the nonlinearity for 5.4~keV x~rays is negligible and at
gain 200 it reaches only 1\%.
RMD LAAPDs have their best resolution at gain 200.
At this gain the nonlinearity for 5.4~keV x~rays is about 5\%, and at
gain 400 reaches 10\%.
The relatively high nonlinearity of the RMD LAAPDs is not an essential
problem for our experiment, because we are interested only in the
intensity variation of the 1.9~keV x~rays as a function of the laser
frequency.
A higher gain achievable with the RMD LAAPDs in feasible x--ray
spectroscopy measurements represents an advantage, especially for an
experiment realized in a high--noise environment.

Our final data taking was performed at typical gains of 400 with a
satisfactory energy resolution.
All 20~RMD LAAPDs were operated for several weeks without any
deterioration.
At the end it was proved that the RMD LAAPDs are suitable for soft
x--ray spectroscopy measurements.

\section{Acknowledgment}
\label{sec:Acknow}

Support is acknowledged from the Swiss National Science Foundation,
the Swiss Academy of Engineering Sciences, from the Portuguese
Foundation for Science and Technology (FCT) Lisbon, and FEDER through
the Project POCTI/FNU/41720/2001, and the program PAI Germaine~de
Sta{\"e}l n$^\circ$07819NH du minist{\`e}re des affaires
{\`e}trang{\'e}res France. The project was in part supported by the US
Department of Energy. Laboratoire Kastler Brossel is an Unit{\'e} Mixte
de Recherche du CNRS n${\circ}$ 8552. Financial support for this
research was received from the Swedish Science Research Councils
(VR).

The authors would like to thank L.~Simons and B.~Leoni for setting up
the cyclotron trap. We also thank the PSI accelerator division, PSI
Hallendienst, PSI and University Fribourg workshops, and other support
groups for their valuable help.

\bibliographystyle{elsart-num}

\end{document}